# Convolutional Neural Networks for Automated Annotation of Cellular Cryo-Electron Tomograms

Muyuan Chen [1,2], Wei Dai [2, #], Stella Y. Sun [2], Darius Jonasch [2], Cynthia Y. He [3], Michael F. Schmid [2], Wah Chiu [2], Steven J. Ludtke [2, *]

[1] Graduate Program in Structural and Computational Biology and Molecular Biophysics, Baylor College of Medicine, Houston, TX 77030

[2] Verna Marrs and McLean Department of Biochemistry and Molecular Biology, Baylor College of Medicine, Houston, TX 77030

[3] Department of Biological Science, Centre for BioImaging Sciences, National University of Singapore, 14 Science Drive 4, Singapore 117543

[#] Current Address: Department of Cell Biology and Neuroscience, Center for Integrative Proteomics Research, Rutgers University, Piscataway, NJ 08854 USA

[*] Corresponding author

# Abstract

Cellular Electron Cryotomography (CryoET) offers the ability to look inside cells and observe macromolecules frozen in action. A primary challenge for this technique is identifying and extracting the molecular components within the crowded cellular environment. We introduce a method using neural networks to dramatically reduce the time and human effort required for subcellular annotation and feature extraction. Subsequent subtomogram classification and averaging yields *in-situ* structures of molecular components of interest.

Cellular Electron Cryotomography (CryoET) is the dominant technique for studying the structure of interacting, dynamic complexes in their native cellular environment at nanometer resolution. While fluorescence imaging techniques can provide high resolution localization of labeled complexes within the cell, they cannot determine the structure of the molecules themselves. X-ray crystallography and single particle CryoEM can study the high resolution structures of macromolecules, but these must first be purified, and cannot be studied *in situ*. In the environment of the cell, complexes are typically transient and dynamic, thus CryoET provides information about cellular processes not attainable by any other current method.

The major limiting factors in cellular CryoET data interpretation include high noise levels, the missing wedge artifact due to experimental geometry (Fig.S2), and the need to study crowded macromolecules that undergo continuous conformational changes[1]. A great deal can be learned by simply annotating the contents of the cell and observing spatial interrelationships among macromolecules. Beyond this, subvolumes can be extracted and aligned to improve resolution by reducing noise and eliminating missing wedge artifacts[2,3]. Unfortunately this is often limited by the extensive conformational and compositional variability within the cell[4,5], requiring classification to achieve more homogeneous populations. Before particles can be extracted and averaged, (macro)molecules of one type must be identified with high fidelity. This task, and the broader task of annotating the contents of the cell, is typically performed by human annotators, and is extremely labor-intensive, requiring as much as one man week for an expert to annotate a typical (4k x 4k x 1k) tomogram. With automated methods now able to produce many cellular tomograms per microscope per day, annotation has become a primary time-limiting factor in the processing pipeline[6].

While algorithms have been developed for automatic segmentation of specific features (for example [7–9]), typically each class of feature has required a separate development effort, and a generalizable algorithm for arbitrary feature recognition has been lacking. We have developed a method based on convolutional neural networks (CNN), which is capable of learning a wide range of possible feature types, and effectively replicates the behavior of a specific expert human annotator. The network requires only minimal human training, and the structure of the network itself is fixed. This method can readily discriminate subtle differences such as double membranes of mitochondria vs. single membranes of other organelles. This single algorithm works well on the major classes of geometrical objects: extended filaments such as tubulin or actin, membranes (curved/planar surfaces), periodic arrays and isolated macromolecules.

Deep neural networks have been broadly applied across many applications in recent years[10]. Past CryoEM applications of neural networks have been limited to simpler methods developed before deep learning, which do not perform well on tomographic data. Among the various deep neural network concepts, deep CNNs are especially useful for pattern recognition in images. While ideally we would develop a single network capable of annotating all known cellular features, the varying noise levels, different artifacts and features in different cell types and large computational requirements makes this impractical at present.

A more tractable approach is, instead, to simplify the problem by training one CNN for each feature to be recognized then merge the results from multiple networks into a single multi-component annotation (Fig.1 and Supplementary movie 1). We have successfully designed a CNN with only a few layers, containing wider than typical kernels (see Fig.S1), which can, with minimal training, successfully identify a wide range of different features across a diverse range of cellular tomograms. The network can be trained quickly, with as few as 10 manually annotated image tiles (64 x 64 pixels) containing the feature of interest and 100 tiles lacking the targeted feature. The CNN-based method we have developed operates on the tomogram slice-by-slice, rather than in 3D, similar to most current manual annotation programs. This approach greatly reduces the complexity of the neural network, improves computational speed, largely avoids the distortions due to the missing wedge artifact[1] (Fig.S2), and still performs extremely well.

Once a CNN has been trained to recognize a certain feature, it can be used to annotate the same feature in other tomograms of the same type of cell under similar imaging conditions without additional training. While difficult to quantitatively assess, we have found that training on one tomogram is generally adequate to successfully annotate all of the tomograms within a particular biological project (see Fig.S4). This enables rapid annotation of large numbers of cells of the same type with minimal human effort, and a reasonable amount of computation.

When multiple scientists are presented with the same tomogram, annotation results are not identical[11,12]. While results are grossly similar in most cases, the annotation of specific voxels can vary substantially among users. Nonetheless, when presented with each other's results most annotators agree that those alternative annotations were also reasonable. That is to say, it is extremely difficult to establish a single "ground truth" in any cellular annotation process. Neural networks make it practical to explore this variability. Rather than having multiple annotators train a single CNN, a CNN can be trained for each annotator for each feature of interest, to produce both a consensus annotation as well as a measure of uncertainty. Given the massive time requirements for manual annotation, this kind of study would never be practical on a large scale using real human annotations, but to the extent that a trained CNN can mimic a specific annotator, this sort of virtual comparison is a practical alternative.

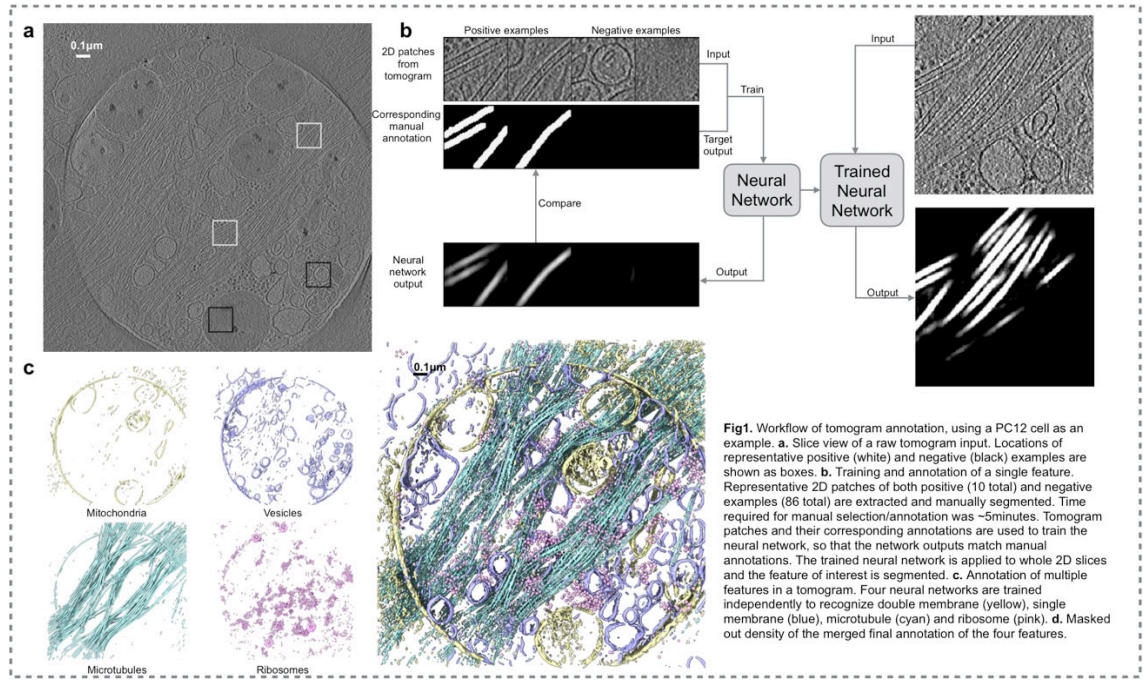

**Fig 1.** Workflow of tomogram annotation, using a PC12 cell as an example. **a.** Slice view of a raw tomogram input. Locations of representative positive (white) and negative (black) examples are shown as boxes. **b.** Training and annotation of a single feature. Representative 2D patches of both positive (10 total) and negative examples (86 total) are extracted and manually segmented. Time required for manual selection/annotation was ~5minutes. Tomogram patches and their corresponding annotations are used to train the neural network, so that the network outputs match manual annotations. The trained neural network is applied to whole 2D slices and the feature of interest is segmented. **c.** Annotation of multiple features in a tomogram. Four neural networks are trained independently to recognize double membrane (yellow), single membrane (blue), microtubule (cyan) and ribosome (pink). **d.** Masked out density of the merged final annotation of the four features.

To test our methodology, we used four distinct cell types: PC12 cells, Human Platelets[13], African Trypanosomes and Cyanobacteria[14]. We targeted multiple subcellular features with various geometries, including bacteriophages, carboxysomes, microtubules, double layer membranes, full and empty vesicles, RuBisCO molecules, and ribosomes. Representative annotations are shown in Fig.2(a-c), Fig.S3 and Supplementary movie 2-5. These datasets were collected on different microscopes, with different defocus and magnification ranges and with or without a phase plate.

After annotation, subtomograms of specific features of interest were extracted from each tomogram, and subtomogram averaging was applied to each set in order to test the usefulness of the annotation. Results are shown in Fig.2(d-f). For instance, the subtomogram average of automatically extracted ribosomes from Trypanosomes resemble a ~40Å resolution version of that structure determined by single particle CryoEM[15]. The structures of microtubules have the familiar features and spacing observed in low resolution structures[16,17], while some have luminal density and others do not[18]. Interestingly, we are able to detect and trace the en face aspect of the thylakoid membrane in cyanobacteria, which is known to have a characteristic pseudo-crystalline array[19] of light harvesting complexes embedded in it, in both the subtomogram patches and their corresponding power spectra, in a single tomogram. These results confirm that the methodology is correctly identifying the subcellular features being trained, with a high level of accuracy. Our example tomograms were collected at low magnification for annotation and qualitative cellular biology, not with subtomogram averaging in mind, and thus the resolution of the averages is limited.

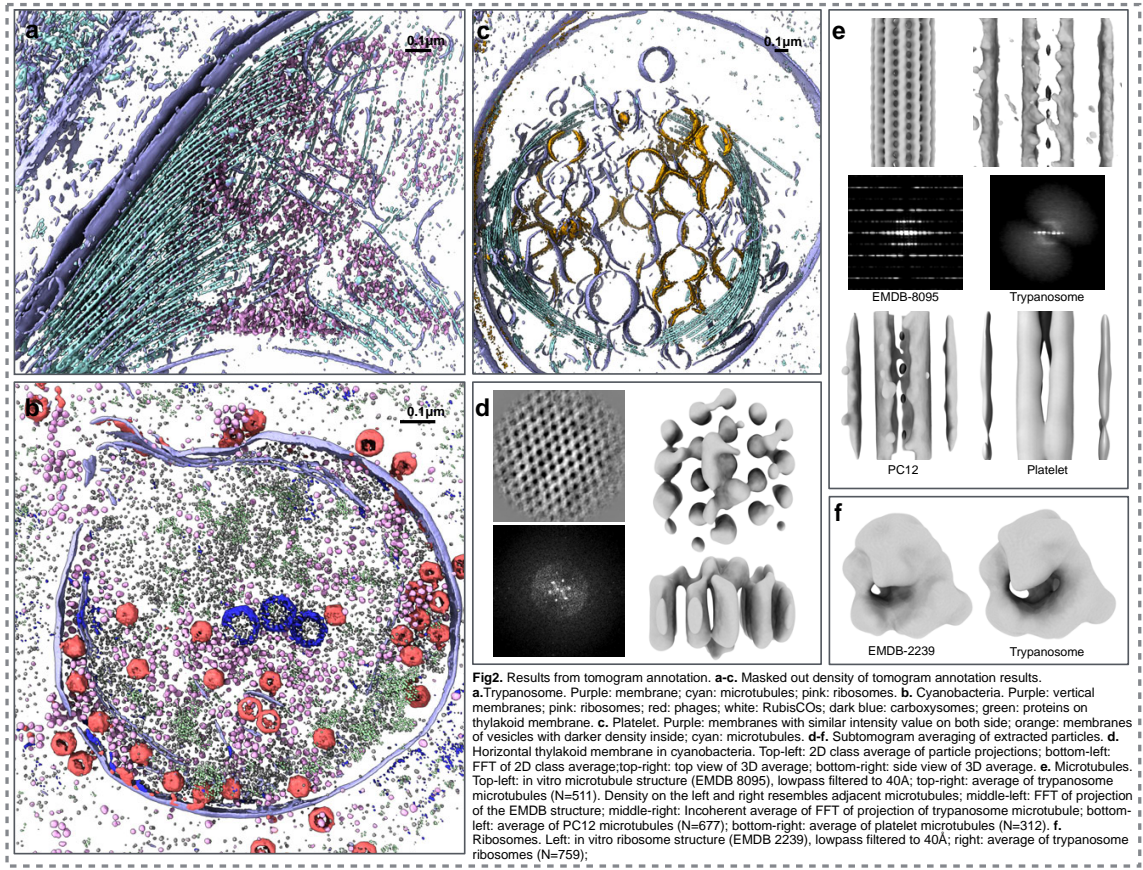

**Fig 2.** Results from tomogram annotation. **a-c.** Masked out density of tomogram annotation results. **a.** Trypanosome. Purple: membrane; cyan: microtubules; pink: ribosomes. **b.** Cyanobacteria. Purple: vertical membranes; pink: ribosomes; red: phages; white: RubisCOs; dark blue: carboxysomes; green: proteins on thylakoid membrane. **c.** Platelet. Purple: membranes with similar intensity value on both side; orange: membranes of vesicles with darker density inside; cyan: microtubules. **d-f.** Subtomogram averaging of extracted particles. **d.** Horizontal thylakoid membrane in cyanobacteria. Top-left: 2D class average of particle projections; bottom-left: FFT of 2D class average;top-right: top view of 3D average; bottom-right: side view of 3D average. **e.** Microtubules. Top-left: in vitro microtubule structure (EMDB 8095), lowpass filtered to 40Å; top-right: average of trypanosome microtubules (N=511). Density on the left and right resembles adjacent microtubules; middle-left: FFT of projection of the EMDB structure; middle-right: Incoherent average of FFT of projection of trypanosome microtubule; bottom-left: average of PC12 microtubules (N=677); bottom-right: average of platelet microtubules (N=312). **f.** Ribosomes. Left: in vitro ribosome structure (EMDB 2239), lowpass filtered to 40Å; right: average of trypanosome ribosomes (N=759);

The set of annotation utilities is freely available as part of the open source package EMAN2.2[20], and includes a graphical interface for training new CNNs as well as applying them to tomograms. A tutorial for using this automated annotation is available online at http://EMAN2.org.

### Acknowledgements:
We gratefully acknowledge support of NIH grants (R01GM080139, P01NS092525, P41GM103832), Ovarian Cancer Research Fund and Singapore Ministry of Education. Molecular graphics and analyses performed with UCSF ChimeraX, developed by the Resource for Biocomputing, Visualization, and Informatics at the University of California, San Francisco (supported by NIGMS P41GM103311).

### Author contributions:
M.C. designed the protocol. W.D, Y.S, C.H provided the test datasets. M.C and D.J. tested and refined the protocol. M.C, W.D, Y.S, M.S., W.C. and S.L wrote the paper and provided suggestions during development.

### Competing financial interests
The authors declare no competing financial interests.

**Reference**:

1. Lučić, V., Rigort, A. & Baumeister, W. Cryo-electron tomography: The challenge of doing structural biology in situ. *J. Cell Biol.* **202,** 407–419 (2013).

2. Galaz-Montoya, J. G. *et al.* Alignment algorithms and per-particle CTF correction for single particle cryo-electron tomography. *J. Struct. Biol.* **194,** 383–94 (2016).

3. Chen, Y., Pfeffer, S., Hrabe, T., Schuller, J. M. & Förster, F. Fast and accurate reference-free alignment of subtomograms. *J. Struct. Biol.* **182,** 235–245 (2013).

4. Asano, S. *et al.* Proteasomes. A molecular census of 26S proteasomes in intact neurons. *Science* **347,** 439–42 (2015).

5. Pfeffer, S., Woellhaf, M. W., Herrmann, J. M. & Förster, F. Organization of the mitochondrial translation machinery studied in situ by cryoelectron tomography. *Nat. Commun.* **6,** 6019 (2015).

6. Ding, H. J., Oikonomou, C. M. & Jensen, G. J. The Caltech Tomography Database and Automatic Processing Pipeline. *J. Struct. Biol.* **192,** 279–86 (2015).

7. Rigort, A. *et al.* Automated segmentation of electron tomograms for a quantitative description of actin filament networks. *J. Struct. Biol.* **177,** 135–44 (2012).

8. Page, C., Hanein, D. & Volkmann, N. Accurate membrane tracing in three-dimensional reconstructions from electron cryotomography data. *Ultramicroscopy* **155,** 20–26 (2015).

9. Frangakis, A. S. *et al.* Identification of macromolecular complexes in cryoelectron tomograms of phantom cells. *Proc. Natl. Acad. Sci.* **99,** 14153–14158 (2002).

10. LeCun, Y., Bengio, Y. & Hinton, G. Deep learning. *Nature* **521,** 436–444 (2015).

11. Garduño, E., Wong-Barnum, M., Volkmann, N. & Ellisman, M. H. Segmentation of electron tomographic data sets using fuzzy set theory principles. *J. Struct. Biol.* **162,** 368–79 (2008).

12. Hecksel, C. W. *et al.* Quantifying Variability of Manual Annotation in Cryo-Electron Tomograms. *Microsc. Microanal.* **22,** 487–496 (2016).

13. Wang, R. *et al.* Electron cryotomography reveals ultrastructure alterations in

platelets from patients with ovarian cancer. *Proc. Natl. Acad. Sci. U. S. A.* **112,** 14266–71 (2015).

14. Dai, W. *et al.* Visualizing virus assembly intermediates inside marine cyanobacteria. *Nature* **502,** 707–710 (2013).

15. Hashem, Y. *et al.* High-resolution cryo-electron microscopy structure of the Trypanosoma brucei ribosome. *Nature* **494,** 385–389 (2013).

16. Asenjo, A. B. *et al.* Structural Model for Tubulin Recognition and Deformation by Kinesin-13 Microtubule Depolymerases. *Cell Rep.* **3,** 759–768 (2013).

17. Koning, R. I. *et al.* Cryo electron tomography of vitrified fibroblasts: microtubule plus ends in situ. *J. Struct. Biol.* **161,** 459–68 (2008).

18. Garvalov, B. K. *et al.* Luminal particles within cellular microtubules. *J. Cell Biol.* **174,** 759–765 (2006).

19. Scheuring, S. Chromatic Adaptation of Photosynthetic Membranes. *Science (80-. ).* **309,** 484–487 (2005).

20. Tang, G. *et al.* EMAN2: An extensible image processing suite for electron microscopy. *J. Struct. Biol.* **157,** 38–46 (2007).

**Figure Captions:**

**Fig 1.** Workflow of tomogram annotation, using a PC12 cell as an example. **a.** Slice view of a raw tomogram input. Locations of representative positive (white) and negative (black) examples are shown as boxes. **b.** Training and annotation of a single feature. Representative 2D patches of both positive (10 total) and negative examples (86 total) are extracted and manually segmented. Time required for manual selection/annotation was ~5minutes. Tomogram patches and their corresponding annotations are used to train the neural network, so that the network outputs match manual annotations. The trained neural network is applied to whole 2D slices and the feature of interest is segmented. **c.** Annotation of multiple features in a tomogram. Four neural networks are trained independently to recognize double membrane (yellow), single membrane (blue), microtubule (cyan) and ribosome (pink). **d.** Masked out density of the merged final annotation of the four features.

**Fig 2.** Results from tomogram annotation. **a-c.** Masked out density of tomogram annotation results. **a.** Trypanosome. Purple: membrane; cyan: microtubules; pink: ribosomes. **b.** Cyanobacteria. Purple: vertical membranes; pink: ribosomes; red: phages; white: RubisCOs; dark blue: carboxysomes; green: proteins on thylakoid membrane. **c.** Platelet. Purple: membranes with similar intensity value on both side; orange: membranes of vesicles with darker density inside; cyan: microtubules. **d-f.** Subtomogram averaging of extracted particles. **d.** Horizontal thylakoid membrane in cyanobacteria. Top-left: 2D class average of particle projections; bottom-left: FFT of 2D class average; top-right: top view of 3D average; bottom-right: side view of 3D average. **e.** Microtubules. Top-left: in vitro microtubule structure (EMDB 8095), lowpass filtered to 40A; top-right: average of trypanosome microtubules (N=511). Density on the left and right resembles adjacent microtubules; middle-left: FFT of projection of the EMDB structure; middle-right: Incoherent average of FFT of projection of trypanosome microtubule; bottom-left: average of PC12 microtubules (N=677); bottom-right: average of platelet microtubules (N=312). **f.** Ribosomes. Left: in vitro ribosome structure (EMDB 2239), lowpass filtered to 40Å; right: average of trypanosome ribosomes (N=759);

# Online Methods:

**Preprocessing**:
Raw tomograms tend to be extremely noisy, and while the CNN can perform filter-like operations, long range filters are more efficiently performed in Fourier space, so this is handled as a preprocessing step. A lowpass filter (typically at 20 Å) and highpass filter (typically at 4000 Å) are performed, the first to reduce noise and the second to reduce ice gradients which may interfere with feature identification. The tomogram is also normalized such that the mean voxel value is 0.0 and the standard deviation is 1.0. Voxels with intensity higher or lower than three times the standard deviation are clamped. These preprocessing steps provide a consistent input range for the CNN.

**Training sets:**
The CNN is trained using 64x64 pixel tiles manually extracted from the tomogram using a specialized graphical tool in EMAN2. Roughly 10 tiles containing the feature of interest must be selected, and then manually annotated with a binary mask using a simple drawing tool. Since this is the only information used to train the CNN it is important to span the range of observed variations for any given feature when selecting tiles to manually annotate. For example, when training for membranes, segmentation performance will be improved if a range of curvatures are included in training. An additional roughly 100 tiles not containing the feature of interest must also be selected as negative examples. Since no annotation is required for these regions, they can be selected very rapidly. The larger number of negative examples is required to cover the diversity of features in the tomogram which are not the feature of interest. The size of 64x64 pixels is fixed in the current design, as a practical trade-off between accuracy, computational efficiency and GPU memory requirements.
Image rotation is not explicitly part of the network structure, so each of the positive training patches is automatically replicated in several random orientations to reduce the number of required manual annotations, as well as compensate for anisotropic artifacts in the image plane. Examples containing the feature of interest are replicated such that the total number of positive and negative examples is roughly equal for training.

**Rationale of neural network design:**
Consider how per-pixel feature recognition would be handled with a traditional neural network. Clearly a pixel cannot be classified strictly based on its own value. To identify the feature it represents, we must also consider some number of surrounding pixels. For example, if we considered a 30x30 pixel region with the pixel under consideration in the center, that would put 900 neurons in the first layer of the neural network, and full connectivity to a second layer would require $900^2$ weights. The number of neurons could be reduced in later layers. However, to give each pixel the advantage of the same number of surrounding pixels, we would have to run a 30x30 pixel tile through the network for each input pixel in the tomogram. For a 4k x 4k x 1k tomogram, assuming a total of ~$5x10^6$ weights in the network and a simple activation function, this would require $10^{17}$-$10^{18}$ floating point operations. That is, annotation of a single tomogram would require days to weeks even with GPU technology.
Similar to the use of FFTs to accelerate image processing operations, the concept of CNNs allows us to handle this problem much more efficiently. In a CNN the concept of a neuron is extended such that a single neuron takes an image as its input and a connection between neurons becomes a convolution operation. These more complicated neurons give the overall network organization a much simpler appearance, since some complexity is hidden within the neurons themselves.

Rotational invariance in the network is handled, as described above, by providing representative tiles in many different orientations, and is one of the major reasons we require 40 neurons. However, the network also needs to locate features irrespective of their translational position within the tile. The use of CNN's in our network design provides effectively performs a translation-independent classification for every pixel in the image simultaneously. That is, it is possible to process an entire slice of a tomogram in a single operation, rather than a tile for every pixel to be classified in the traditional approach.

The concepts underlying deep neural networks is that by combining many layers of very simple perceptron units, it is possible to learn arbitrarily complex features in the data. In addition to dramatically simplifying the network training process, the use of many layers permits arbitrary nonlinear functions to be represented despite the use of a simple ReLU activation function[1]. Many new techniques have been developed around this concept[2–4], improving on the convergence rate and robustness of classical neural network designs, such as the CNN, enabling training of large networks to solve problems which were previously intractable for neural network technology.

In typical use-cases, pretrained neural networks are provided to users, and the users simply apply the existing network to their data[5]. The training process is generally extremely complicated and requires both a skilled programmer and significant computational resources. However, in the current application, pretraining of the network is not feasible, due the lack of appropriate examples spanning the diverse potential user data and features of interest.

Another challenge of biological feature recognition is that biological features cover a wide range of scales. The pooling technique used in convolutional neural network design is capable of spanning such scale differences, but spatial localization precision is sacrificed during this process. Although it is possible to perform unpooling and train interpolation kernels at the end of the network[6], or simply interpolate by applying the neural network on multiple shifted images, those methods make the training process much slower and more difficult to achieve convergence. Given our inability to provide a pretrained network, we needed a design that could be trained reliably in an automatic fashion, so end-users could make use of the system with no knowledge of neural networks. Instead of a single large and deep neural network that performs the entire classification process, we use a set of smaller independent networks, each of which recognizes only a single feature. These smaller networks are then merged to produce the overall classification result. In addition, to reduce the need for max-pooling and permit a shallower structure, we use 15x15 pixel kernels, larger than typical in CNN image processing[7]. This larger kernel size permits us to use only a single 2x2 max pooling layer and is still able to discern features requiring fine detail, such as double layer vs. single layer membranes. The number of neurons was selected through extensive testing with real data, to be as small as possible while remaining accurate.

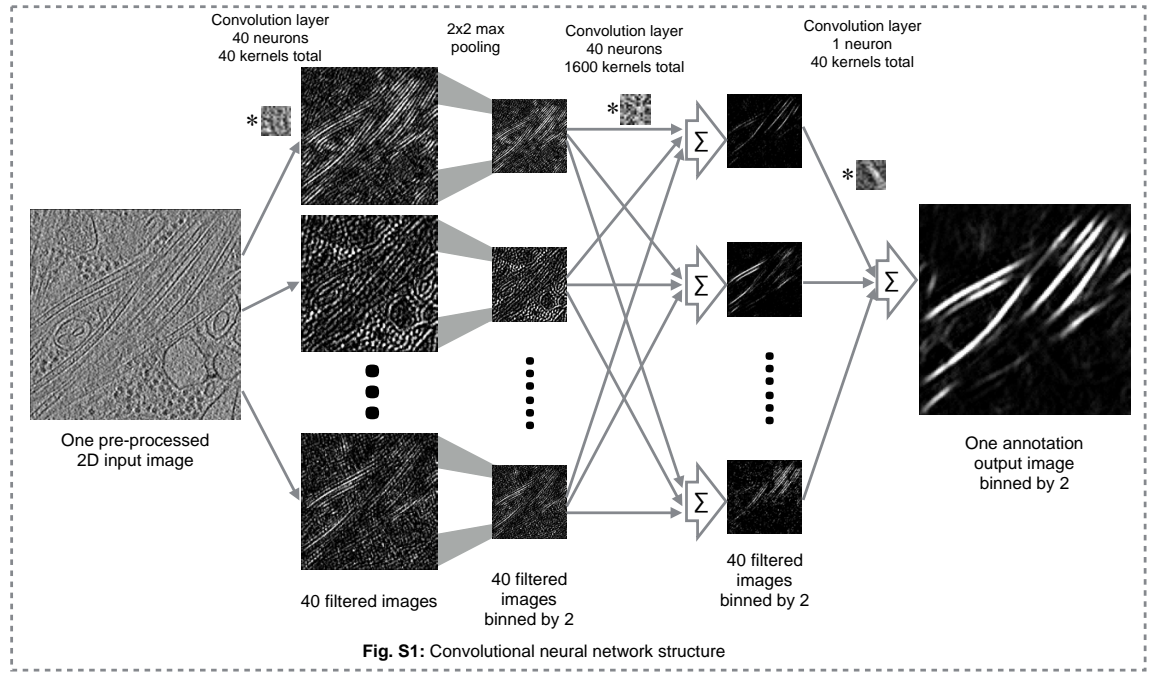

**Fig. S1**: Convolutional neural network structure

**Neural network structure:**
The 4-layer CNN we settled on was used in all of our examples (Fig.S1). The first layer is a convolutional layer, containing 40 neurons, each of which has a 2D kernel with 15x15 pixels and a scalar offset value. In the input to each first layer neuron, particles are filtered by a 2D kernel. A linear offset and a ReLU activation function are applied to the neuron output. The activation function is what permits nonlinear behavior in the network, though the degree of nonlinearity is limited by the small number of layers we are using. Each first layer neuron thus outputs a 64x64 tile. The second layer performs max pooling, which outputs the maximum value in each 2x2 pixel square in the input tiles. This downsamples each tile to 32x32. The second layer is then fully connected to the 40-neuron third layer, again with independent 15x15 convolution kernels for each of the 1600 connections. Although the size of the kernel is the same as in the first layer, thanks to the max pooling, the kernel in this layer can cover features of a larger scale, up to 30x30 pixels, roughly ½ of the training tile size. The fourth and final layer consists of only one neuron producing the final single 32x32 output tile. Finally, the filtered results are summed and a linear offset is applied. To match the results of a manual annotation and expedite convergence, a specialized ReLU activation function ( $y=\min(1,x)$ ) is used.

**Hyper-parameter selection and Neural network training:**
Before training, all the kernels in the neural network are initialized using a uniform distribution of near-zero values, and the offsets are initialized to zero. Log squared residual ( $\log((y-y')^2)$ ) between the neural network output and the manual annotation is used as the loss function. Since there is a pooling layer in the network, the manual annotation is shrunk by 2 to match the network output. A L1 weight decay of $10^{-5}$ is used for regularization of the training process. No significant overfitting is observed, likely because the high noise level in the CryoET images also serves as a strong regularization factor. To optimize the kernels, we use stochastic gradient descent with a batch size of 20. By default, the neural network is trained for 20 iterations. The learning rate is set to 0.01 in the first iteration and decreased by 10% after each iteration. The training process can be performed on either a GPU or in parallel on multiple CPUs (~10x slower on our testing machine). Training each feature typically takes under 10 minutes on a current generation GPU, and the resulting network can be used for any tomogram of the same cell-type collected under similar conditions. A workstation with 96GB of RAM, 2x Intel X5675 processors for a total of 12 compute cores, and an Nvidia GTX 1080 GPU was used for all testing.

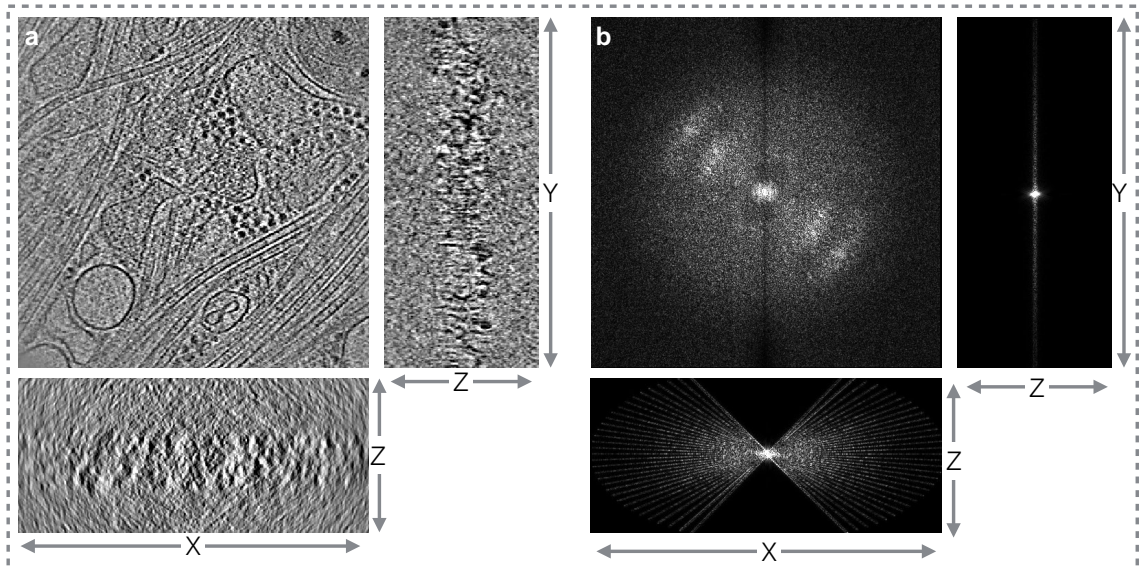

**Fig. S2:** Example of the impact of the missing wedge artifact in CryoET. This demonstrates why tomograms are typically annotated using X-Y slices. **a.** Slice view of a tomogram in the X-Y, Y-Z and X-Z plane. **b.** Fourier Transform of projection of the same tomogram in X-Y, Y-Z and X-Z plane.

**Applying trained CNNs to tomograms:**
Since the neural network is convolutional, we simply filter the full-sized pre-filtered tomogram slices with the trained kernels in the correct order to generate the output. Unlike the training process, the CNN is applied to entire (typically 4k x 4k) tomogram slices. The network is applied to the images by propagating the image as described above in network design. Practical implementation involves simple matrix operations combined with FFTs for the convolution operations. The final output tile is unbinned by 2 to match the dimensions of the input tile. Each voxel in the assembled density map is related to the likelihood that that voxel corresponds to the feature used to train the network. While the networks are trained on the GPU, due to memory limitations and the large number of kernels, applying the trained network is currently done on CPUs, where it can be efficiently parallelized assuming sufficient RAM is available. For reference, annotation of one feature on a 4096x4096x1024 tomogram requires ~20 hours on the test computer. Note that the full 4Kx4K tomogram is only need when annotating small features that are only visible in the unbinned tomogram. Practically, all features in the test datasets shown in this paper were annotated on tomogram binned by 2~8. As the speed of CNN application scales linearly with the number of voxels in the tomogram, it only takes ~2.5 hours to annotate the same tomogram binned by 2. However, there is still room for significant optimization of the code through reorganization of mathematical operations, without altering the network structure.

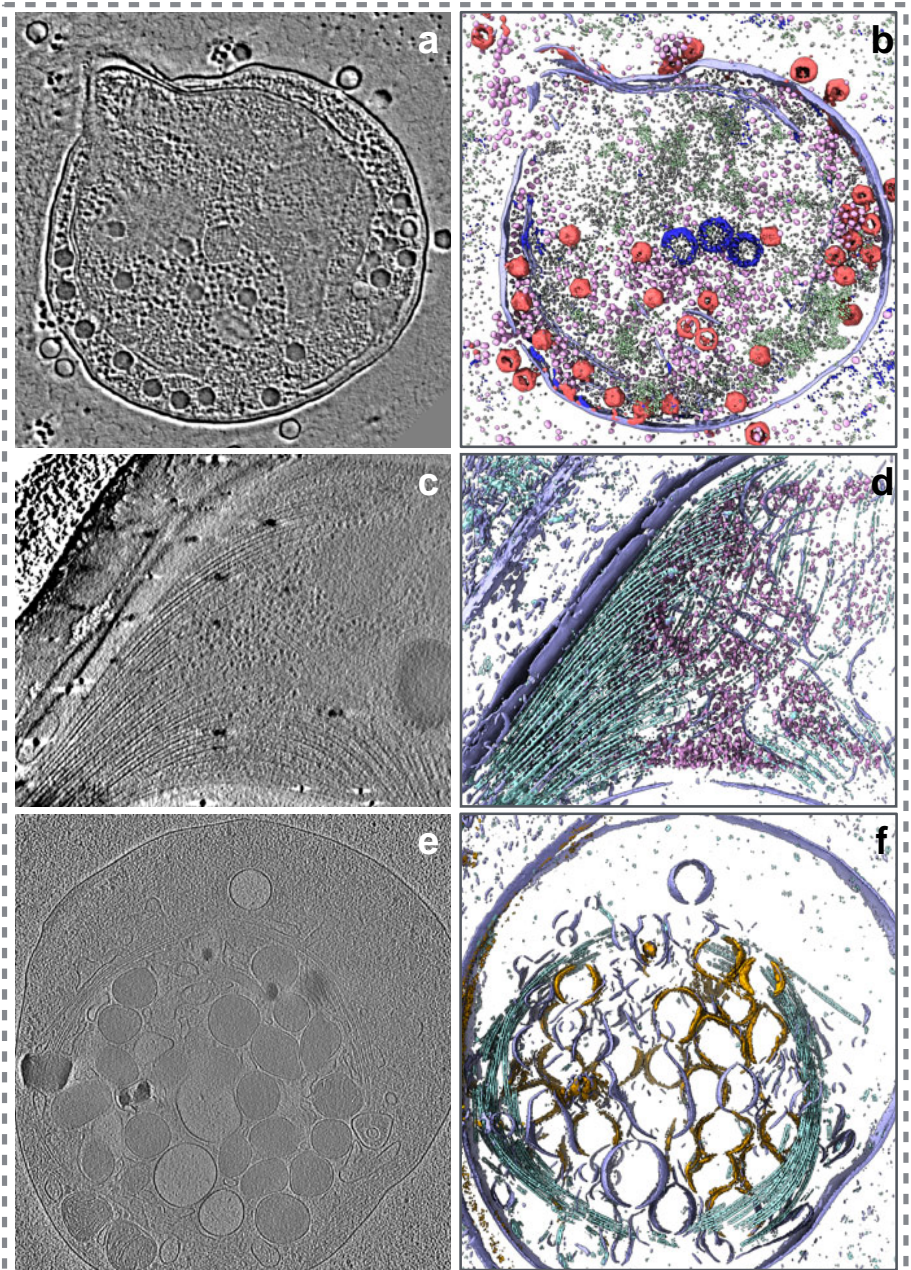

**Fig.S3:** Comparison of a single raw tomogram slice with each corresponding 3-D automatic annotation for each tomogram in **Fig.2**.

**Post processing and merging features:**
After applying a single CNN to a tomogram, the output needs to be normalized so the results from the set of trained networks are comparable. This is done by scaling the output of the neural network annotation so that the mean value on manually annotated regions in positive pixels in the training set is 1. After normalization, annotation results from multiple CNNs can be merged by simply identifying which CNN had the highest value for each voxel. We also use 1 as a threshold value for isosurface display as well as for particle identification/extraction for subtomogram averaging.

## Particle extraction and averaging:

While annotation alone is sufficient for certain types of cellular tomogram interpretation, subtomogram extraction and averaging remains a primary purpose for detailed annotation. To extract discrete objects like ribosomes or other macromolecules, we begin by identifying all connected voxels annotated as being the same feature. For each connected region, the maxima position in the annotation is used as the particle location. For continuous features like microtubules, we randomly seed points on the annotation output and use these points as box coordinates for particle extraction. In both cases, EMAN2 2D classification[8] is performed on a Z-axis projection of the particles in order to help identify and remove bad particles, in a similar fashion to single particle analysis. 3D alignment and averaging are performed using EMAN2 single particle tomography utilities[9].

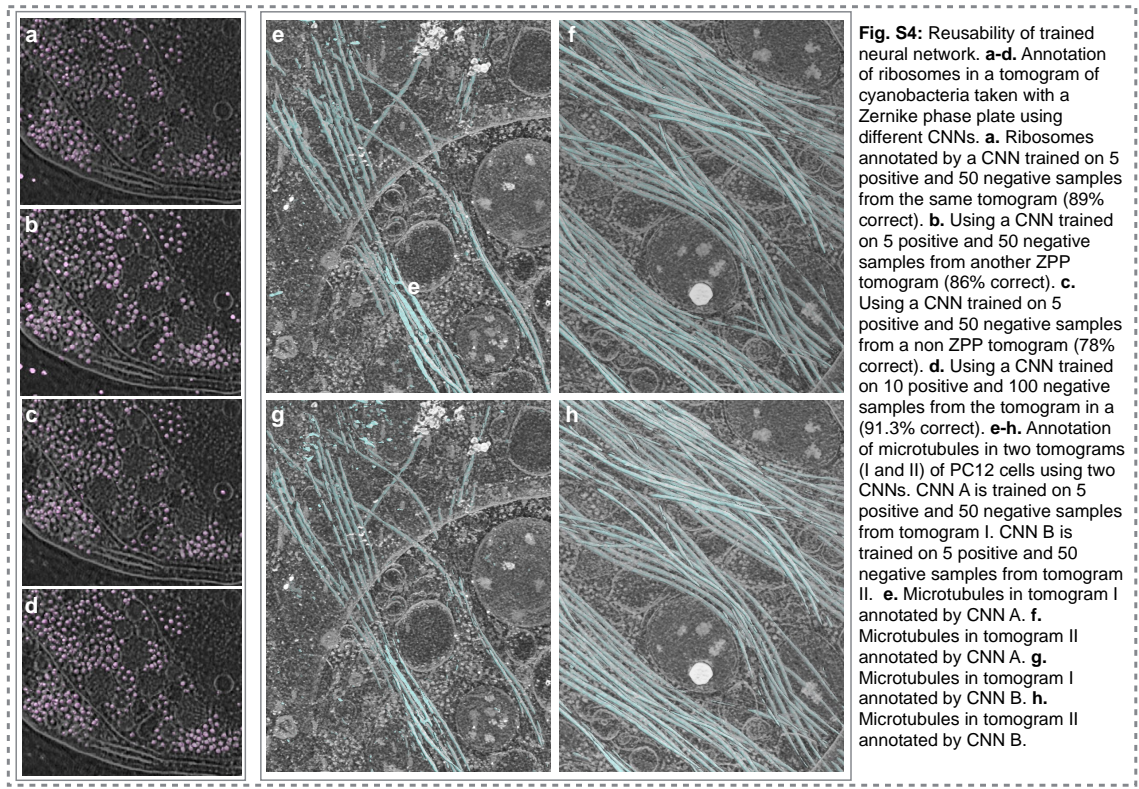

**Fig. S4:** Reusability of trained neural network. **a-d.** Annotation of ribosomes in a tomogram of cyanobacteria taken with a Zernike phase plate using different CNNs. **a.** Ribosomes annotated by a CNN trained on 5 positive and 50 negative samples from the same tomogram (89% correct). **b.** Using a CNN trained on 5 positive and 50 negative samples from another ZPP tomogram (86% correct). **c.** Using a CNN trained on 5 positive and 50 negative samples from a non ZPP tomogram (78% correct). **d.** Using a CNN trained on 10 positive and 100 negative samples from the tomogram in a (91.3% correct). **e-h.** Annotation of microtubules in two tomograms (I and II) of PC12 cells using two CNNs. CNN A is trained on 5 positive and 50 negative samples from tomogram I. CNN B is trained on 5 positive and 50 negative samples from tomogram II. **e.** Microtubules in tomogram I annotated by CNN A. **f.** Microtubules in tomogram II annotated by CNN A. **g.** Microtubules in tomogram I annotated by CNN B. **h.** Microtubules in tomogram II annotated by CNN B.

## Reusability:

Once a neural network is trained to recognize a feature in one tomogram, within some limits, it can be used to annotate the same feature in other tomograms. These tomograms should have the same voxel size and have been pre-processed in the same way. For some universal features like the membranes, it would be possible to apply a trained neural network on a completely different cell type, but for most other features, it is strongly recommended that the neural network be trained on a tomogram of the same cell type under similar imaging conditions. In practice, we have found performance of the neural network to be robust to reasonable differences in defocus or signal to noise ratio. These statements are difficult to quantify, however, since different types of features have different sensitivity to such factors.

To perform a rough quantification of network reusability, we segmented ribosomes from a single cyanobacteria tomogram collected using a Zernike phase plate[10] using 4 different CNNs, each trained in a different way (Fig.S4). We used each CNN to identify ribosomes, and then manually identified which of the putative ribosomes appeared to be something else. We tested only for false positives, not false negatives. While this is not a robust test, since we lack ground truth, we believe it does at least give a general idea about reusability of CNNs. All four CNNs identified ~800-900 ribosomes in the test tomogram. It should be noted that this test used a single CNN for classification, whereas in a normal use-case, multiple CNNs would be competing for each voxel, improving classification accuracy. Given this, the achieved accuracy levels are actually quite high. In the first test, we trained a CNN using the test tomogram itself, with only 5 positive samples and 50 negative samples. We estimate that 89% of the particles were correctly identified. In the second test, a tomogram from the same set, also using the Zernike phase plate was used for training, and the network achieved an estimated 86% accuracy. The third CNN was trained using a tomogram collected with conventional imaging (no phase plate), with an estimated accuracy of 78%. We note that in this test, most of the false positives were due to cut-on artifacts in the phase plate tomogram. The final neural network was again trained on the test tomogram, but this time with 10 positive examples and 100 negative examples. This improved performance to 91%, somewhat better than the first test with fewer training tiles. While ribosomes are somewhat larger and denser than most other macromolecules in the cell, even something somewhat smaller, like a free RuBisCO could potentially be misidentified if a competitive network were not being applied for this feature. For this reason, we consider 75%+ performance with a single network to be extremely good. We also show a cross-test on microtubules in Fig.S4.

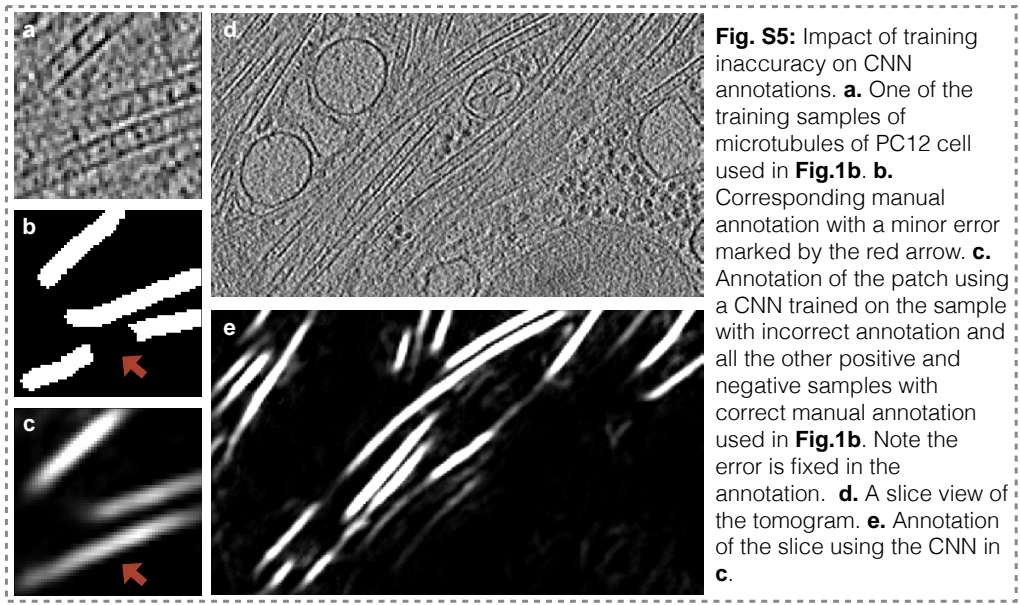

**Fig. S5:** Impact of training inaccuracy on CNN annotations. **a.** One of the training samples of microtubules of PC12 cell used in **Fig.1b**. **b.** Corresponding manual annotation with a minor error marked by the red arrow. **c.** Annotation of the patch using a CNN trained on the sample with incorrect annotation and all the other positive and negative samples with correct manual annotation used in **Fig.1b**. Note the error is fixed in the annotation. **d.** A slice view of the tomogram. **e.** Annotation of the slice using the CNN in **c**.

**Limitations:**
As a machine learning model, the only knowledge input to the CNN is from the provided training set. Although the network is robust to minor variations of the target feature and manual annotation in the training set (Fig.S5), its performance is unpredictable on features which have not been specifically trained for in any of the CNNs. Some non-biological features, such as carbon edges and gold fiducials have much higher contrast than the biological material, and may cause locally unstable behavior of the neural network (Fig.1(c-d)). While misidentification of these features is not generally a problem in display and subtomogram averaging (they can easily be marked as outliers and get removed in the average), it is possible to train a CNN to only recognize these non-biological features and effectively remove them from the annotation output (Fig.S6). In addition, due to the design of the neural network, the maximum scale range of detectable feature is 30 pixels. That is to say, the largest attribute used to recognize a feature of interest can be at most 30 times larger than the smallest. For example, it is readily possible to discriminate between membranes associated with "darker vesicles" from the membrane of light-colored vesicles in platelets (Fig.2c). However, annotating regions inside the "darker vesicles" works poorly. This is because, to annotate regions inside those vesicles, the two aspects of this feature are "high intensity value" and "enclosed by membrane", and at the center of a large vesicle, the distance to the nearest membrane may be more than 30 times larger the thickness of the membrane, so the neural network will have difficulty recognizing the two aspects at the same time, often leaving the center of vesicles empty. This does not prevent the entire set of CNNs from accurately discriminating both large and small objects, since the scale can be set differently for each CNN by pre-scaling the input tomogram.

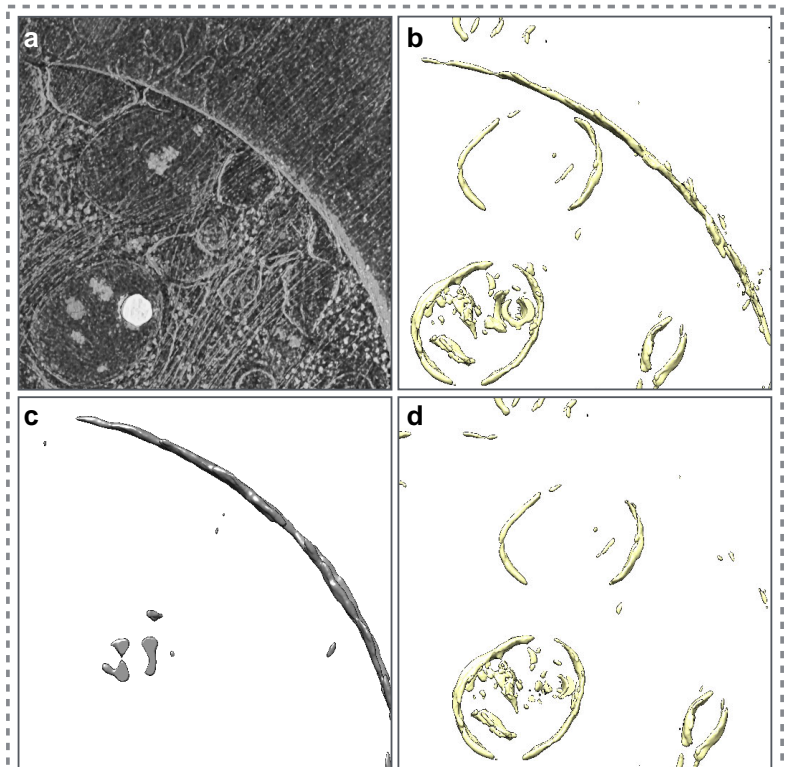

**Fig. S6:** Removal of high contrast artifact. **a.** Volume rendering of a region of the tomogram shown in **Fig 1**. **b.** Annotation of double membrane in the tomogram using the same CNN used in **Fig 1**. Note the carbon edge is falsely recognized as double membrane. **c.** Automatic annotation of carbon edge using a CNN. **d.** Removal of carbon edge from double membrane annotation in **b.**

**Data source:**
PC12 cells were obtained from Dr. Leslie Thompson from UC Irvine[11] and the tomograms were collected on a JEM2100 microscope with a CCD camera. Platelet and cyanobacteria tomograms used in this paper are from previously published dataset[12,13]. Trypanosome cells used in this paper come from procyclic form 29.13 cell line engineered for tetracycline-inducible expression[14]. The tomogram was collected on a JEM2200FS microscope with a DirectElectron DE12 detector.

**Code availability :**
The set of annotation utilities is freely available as part of the open source package EMAN2.2. Both binary download and source code can be found online through http://EMAN2.org.

**Data availability:**
Subtomogram averaging results and one of the full cell annotations are deposited in EMDatabank.
EMD-8589: Subtomogram average of microtubules from Trypanosoma brucei
EMD-8590: Subtomogram average of ribosome from Trypanosoma brucei
EMD-8591: Subtomogram average of protein complexes on the thylakoid membrane in Cyanobacteria
EMD-8592: Subtomogram average of microtubules from PC12 cell
EMD-8593: Subtomogram average of microtubules from Human platelet
EMD-8594: Automated tomogram annotation of PC12 cell

**Supplementary Table: Example details**

| Dataset name | Imaging condition | Å/pix | Training set size (positive/negative) |
| --- | --- | --- | --- |
| PC12 | JEM2100, CCD | 7.2 | Membrane (10/55); Microtubule (10/86); Ribosome (7/67); Mitochondria (10/71) |
| Platelet | JEM2200FS, CCD | 12.0 | Empty vesicle (9/87); Filled vesicle (7/101); Microtubule (8/140) |
| Cyanobacteria | JEM2200, CCD, Phase plate | 5.2 | Membrane (16/88); Ribosome (7/101); RuBisCO (9/107); Thylakoid membrane (8/107); Carboxysome (6/107); Phage (10/101) |
| Trypanosome | JEM2200, DE12 | 5.3 | Membrane (11/100); Microtubule (8/147); Ribosome (5/125) |

**Reference:**

1.  Nair, V. & Hinton, G. E. Rectified Linear Units Improve Restricted Boltzmann

Machines. *Proc. 27th Int. Conf. Mach. Learn.* 807–814 (2010). doi:10.1.1.165.6419

2. Vincent, P., Larochelle, H., Bengio, Y. & Manzagol, P.-A. Extracting and composing robust features with denoising autoencoders. *Proc. 25th Int. Conf. Mach. Learn. - ICML '08* 1096–1103 (2008). doi:10.1145/1390156.1390294

3. Hinton, G. E., Srivastava, N., Krizhevsky, A., Sutskever, I. & Salakhutdinov, R. R. Improving neural networks by preventing co-adaptation of feature detectors. (2012). at <http://arxiv.org/abs/1207.0580>

4. Dieleman, S., Willett, K. W. & Dambre, J. Rotation-invariant convolutional neural networks for galaxy morphology prediction. *Mon. Not. R. Astron. Soc.* **450,** 1441–1459 (2015).

5. Zhou, J. & Troyanskaya, O. G. Predicting effects of noncoding variants with deep learning–based sequence model. *Nat. Methods* **12,** 931–934 (2015).

6. Noh, H., Hong, S. & Han, B. Learning Deconvolution Network for Semantic Segmentation. (2015). at <http://arxiv.org/abs/1505.04366>

7. Krizhevsky, A., Sutskever, I. & Hinton, G. E. ImageNet Classification with Deep Convolutional Neural Networks. *Adv. Neural Inf. Process. Syst.* 1–9 (2012). doi:http://dx.doi.org/10.1016/j.protcy.2014.09.007

8. Tang, G. *et al.* EMAN2: An extensible image processing suite for electron microscopy. *J. Struct. Biol.* **157,** 38–46 (2007).

9. Galaz-Montoya, J. G. *et al.* Alignment algorithms and per-particle CTF correction for single particle cryo-electron tomography. *J. Struct. Biol.* **194,** 383–94 (2016).

10. Dai, W. *et al.* Zernike phase-contrast electron cryotomography applied to marine cyanobacteria infected with cyanophages. *Nat. Protoc.* **9,** 2630–2642 (2014).

11. Apostol, B. L. *et al.* A cell-based assay for aggregation inhibitors as therapeutics of polyglutamine-repeat disease and validation in Drosophila. *Proc. Natl. Acad. Sci.* **100,** 5950–5955 (2003).

12. Wang, R. *et al.* Electron cryotomography reveals ultrastructure alterations in platelets from patients with ovarian cancer. *Proc. Natl. Acad. Sci. U. S. A.* **112,** 14266–71 (2015).

13. Dai, W. *et al.* Visualizing virus assembly intermediates inside marine cyanobacteria. *Nature* **502,** 707–710 (2013).

14. Wirtz, E., Leal, S., Ochatt, C. & Cross, G. A. A tightly regulated inducible expression system for conditional gene knock-outs and dominant-negative genetics in Trypanosoma brucei. *Mol. Biochem. Parasitol.* **99,** 89–101 (1999).

Supplementary Figure Captions:

**Fig. S1:** Convolutional neural network structure

**Fig. S2:** Example of the impact of the missing wedge artifact in CryoET. This demonstrates why tomograms are typically annotated using X-Y slices. **a.** Slice view of a tomogram in the X-Y, Y-Z and X-Z plane. **b.** Fourier Transform of projection of the same tomogram in X-Y, Y-Z and X-Z plane.

**Fig. S3:** Comparison of raw tomogram slices and corresponding automatic annotation in **Fig 2**.

**Fig. S4:** Reusability of trained neural network. **a-d.** Annotation of ribosomes in a tomogram of cyanobacteria taken with a Zernike phase plate using different CNNs. **a.** Ribosomes annotated by a CNN trained on 5 positive and 50 negative samples from the same tomogram (89% correct). **b.** Using a CNN trained on 5 positive and 50 negative samples from another ZPP tomogram (86% correct). **c.** Using a CNN trained on 5 positive and 50 negative samples from a non ZPP tomogram (78% correct). **d.** Using a CNN trained on 10 positive and 100 negative samples from the tomogram in a (91.3% correct). **e-h.** Annotation of microtubules in two tomograms (I and II) of PC12 cells using two CNNs. CNN A is trained on 5 positive and 50 negative samples from tomogram I. CNN B is trained on 5 positive and 50 negative samples from tomogram II. **e.** Microtubules in tomogram I annotated by CNN A. **f.** Microtubules in tomogram II annotated by CNN A. **g.** Microtubules in tomogram I annotated by CNN B. **h.** Microtubules in tomogram II annotated by CNN B.

**Fig. S5:** Impact of training inaccuracy on CNN annotations. **a.** One of the training samples of microtubules of PC12 cell used in **Fig.1b**. **b.** Corresponding manual annotation with a minor error marked by the red arrow. **c.** Annotation of the patch using a CNN trained on the sample with incorrect annotation and all the other positive and negative samples with correct manual annotation used in **Fig.1b**. Note the error is fixed in the annotation. **d.** A slice view of the tomogram. **e.** Annotation of the slice using the CNN in **c**.

**Fig. S6:** Removal of high contrast artifact. **a.** Volume rendering of a region of the tomogram shown in **Fig 1**. **b.** Annotation of double membrane in the tomogram using the same CNN used in **Fig 1**. Note the carbon edge is falsely recognized as double membrane. **c.** Automatic annotation of carbon edge using a CNN. **d.** Removal of carbon edge from double membrane annotation in **b**.